\documentclass[10pt,reqno,a4paper]{amsart}
\usepackage{mathrsfs}
\usepackage{bbm}
\usepackage[numbers,sort&compress]{natbib}

\newtheorem{theorem}{Theorem}
\newtheorem{lemma}[theorem]{Lemma}
\newtheorem{proposition}[theorem]{Proposition}
\newtheorem{observation}[theorem]{Observation}

\newcommand{\bra}[1]{{\langle{#1}\rvert}}
\newcommand{\ket}[1]{{\lvert{#1}\rangle}}
\newcommand{\proj}[1]{{\ket{#1}\!\bra{#1}}}
\newcommand{\ignore}[1]{{\lfloor{#1}\rfloor}}
\newcommand{\eval}[1]{{\langle{#1}\rangle}}

\DeclareMathOperator{\tr}{tr}
\providecommand{\openone}{{\mathbbm 1}}

\newcommand{\CC}{{\mathbf C}}
\newcommand{\RR}{{\mathbf R}}
\newcommand{\mcost}{{\mathrm M}}
\newcommand{\state}[1]{{\mathnormal{S}_{#1}}}
\newcommand{\sqset}{{\mathscr Q}}
\newcommand{\automaton}{{\mathcal A}}
\newcommand{\taurus}{{\automaton_3}}
\newcommand{\gemini}{{\automaton_4}}

\begin{document}
\title{Memory cost of quantum contextuality}

\author[M Kleinmann, O G{\"u}hne, J R Portillo, J-{\AA} Larsson,
A Cabello]{Matthias Kleinmann$^{1,2}$, Otfried G{\"u}hne$^{1,2,3}$, Jos\'{e} R 
Portillo$^4$,
Jan-{\AA}ke Larsson$^5$ and Ad\'{a}n Cabello$^{6,7}$}
\thanks{$^1$ Institut f\"ur Quantenoptik und Quanteninformation,
 \"Osterreichische Akademie der Wissenschaften,
 Technikerstr.~21A, A-6020 Innsbruck, Austria}
\thanks{$^2$ Universit\"at Siegen, Fachbereich Physik,
Walter-Flex-Stra\ss{}e 3, D-57068 Siegen, Germany}
\thanks{$^3$ Institut f\"ur Theoretische Physik,
 Universit\"at Innsbruck,
 Technikerstr.~25, A-6020 Innsbruck, Austria}
\thanks{$^4$ Departamento de Mathem\'atica Aplicada I,
 Universidad de Sevilla,
 E-41012 Sevilla, Spain}
\thanks{$^5$ Institutionen f\"or Systemteknik och Matematiska Institutionen,
 Link\"opings Universitet,
 SE-581 83 Link\"oping, Sweden}
\thanks{$^6$ Departamento de F\'{\i}sica Aplicada II,
 Universidad de Sevilla,
 E-41012 Sevilla, Spain}
\thanks{$^7$ Department of Physics, Stockholm University,
 S-10691 Stockholm, Sweden}

\begin{abstract}
The simulation of quantum effects requires certain classical resources, and 
 quantifying them is an important step in order to characterize the difference 
 between quantum and classical physics.
For a simulation of the phenomenon of state-independent quantum contextuality, 
 we show that the minimal amount of memory used by the simulation is the 
 critical resource.
We derive optimal simulation strategies for important cases and prove that 
 reproducing the results of sequential measurements on a two-qubit system 
 requires more memory than the information carrying capacity of the system.
\end{abstract}

\maketitle

\section{Introduction}
According to quantum mechanics, the result of a measurement may depend on which 
 other compatible observables are measured simultaneously 
 \cite{Specker:1960DIE, Kochen:1967JMM, Bell:1966RMP}.
This property is called contextuality and is in contrast to classical physics, 
 where the answer to a single question does not depend on which other 
 compatible questions are asked at the same time.

Contextuality can be seen as complementary to the well known nonlocality of 
 distributed quantum systems \cite{Mermin:1990PRL}.
Both phenomena can be used for information processing tasks, albeit the 
 applications of contextuality are far less explored \cite{Heywood:1983FPH, 
 BechmannP:2000PRL, Spekkens:2008PRL, Cabello:2010PRL, Aharon:2008PRA, 
 Svozil:2009PRA, Svozil:2009XXX, Horodecki:2010XXX}.
Although contextuality and nonlocality can be considered as signatures of 
 nonclassicality, they can be simulated by classical models \cite{Bell:1966RMP, 
 Bell:1982FOP, LaCour:2009PRA, Khrennikov:2009}.
However, while nonlocal classical models violate a fundamental physical 
 principle (the bounded speed of information), it is not clear whether 
 contextual classical models violate any fundamental principle.
Moreover, while the resources needed in order to imitate quantum nonlocality 
 have been extensively investigated \cite{Toner:2003PRL, Pironio:2003PRA, 
 Buhrman:2010RMP}, there is no similar knowledge about the resources needed to 
 simulate quantum contextuality.

In any model which exhibits contextuality in sequential measurements, the 
 system will eventually attain different internal states during certain 
 measurement sequences.
These states can be considered as \emph{memory} --- a model attaining the 
 minimal number of states is then memory-optimal and defines the memory cost.
In this paper we investigate the memory cost as the critical resource in a 
 classical simulation of quantum contextuality and we construct memory-optimal 
 models for relevant cases.
The amount of required memory increases as we consider more and more 
 contextuality constraints.
This can be used to quantify contextuality in a given quantum setting.
We show that certain scenarios breach the amount of two bits needed for the 
 simulation of two qubits.
This demonstrates that the memory needed to simulate only a small set of 
 measurements on a quantum system may exceed the information that can be 
 transmitted using this system (given by the Holevo bound 
 \cite{Holevo:1973PIT}) --- a similar effect was observed so far only for the 
 classical simulation of a \emph{unitary} evolution \cite{Galvao:2003PRL}.

\section{Scenario}
We focus on the following set of two-qubit observables, also known as the 
 Peres-Mermin (PM) square \cite{Peres:1990PLA, Mermin:1990PRL},
\begin{equation}
\left[\begin{matrix}
 A&B&C\\
 a&b&c\\
 \alpha&\beta&\gamma
\end{matrix}\right]
=
\left[\begin{matrix}
 \sigma_z \otimes \openone&
\openone \otimes \sigma_z&
\sigma_z \otimes \sigma_z\\
 \openone \otimes \sigma_x&
 \sigma_x \otimes \openone&
 \sigma_x \otimes \sigma_x\\
 \sigma_z \otimes \sigma_x&
 \sigma_x \otimes \sigma_z&
 \sigma_y \otimes \sigma_y
\end{matrix}\right],
\end{equation}
 where $\sigma_x$, $\sigma_y$, and $\sigma_z$ denote the Pauli operators.
The square is constructed such that the observables within each row and column 
 commute and are hence compatible, and the product of the operators in a row or 
 column yields $\openone$, except for the last column where it yields 
 $-\openone$.
Thus the product of the measurement results for each row and column will be 
 $+1$ except in the third column where it will be $-1$.
In contrast, for a noncontextual model the measurement result for each 
 observable must not depend on whether the observable is measured in the column 
 or row context.
Hence the number of rows and columns yielding a product of $-1$ is always even, 
 as any observable appears twice.

Similar to the Bell inequalities for local models, any noncontextual model 
 satisfies the inequality
\begin{equation}
 \eval\chi \equiv \eval{ABC}+ \eval{abc}+ \eval{\alpha\beta\gamma}+ 
 \eval{Aa\alpha}+ \eval{Bb\beta}- \eval{Cc\gamma} \le 4,
\end{equation}
while for perfect observables quantum mechanics (QM) predicts $\eval\chi = 6$ 
 \cite{Cabello:2008PRL}.
Here, the term $\eval{ABC}$ denotes the average value of the product of the 
 outcomes of $A$, $B$, and $C$, if these observables are measured 
 simultaneously or in sequence on the same quantum system.
The violation is independent of the quantum state, which emphasizes that the 
 phenomenon is a property of the set of observables rather than of a particular 
 quantum state.

Recently, this inequality has been experimentally tested using trapped ions 
 \cite{Kirchmair:2009NAT}, photons \cite{Amselem:2009PRL}, and nuclear magnetic 
 resonance systems \cite{Moussa:2010PRL}.
The results show a good agreement with the quantum predictions.
In these experiments, the observables are measured in a sequential manner.
Since the observed results cannot be explained by a model using only 
 preassigned values, the system necessarily attains different states during 
 some particular sequences, i.e., the system memorizes previous events.
(Note that also in QM the system attains different states during the 
 measurement sequences.)
This leads to our central question:
\emph{How much memory is required in order to simulate quantum contextuality?}

\section{A first model}
Before we formulate the previous question more precisely, let us provide an 
 example of a model that simulates the contextuality in the PM square.
We assume that the system can only attain {3} different physical states $\state 
 1$, $\state 2$, and $\state 3$  (e.g.\ discrete points in phase-space).
Let us associate a table to each state via
\begin{equation}
\state 1\!:\! \left[\begin{matrix}
     +&+&(+,2)\\
     +&+&(+,3)\\
     +&+&+
    \end{matrix}\right]\!,\;\;
\state 2\!:\! \left[\begin{matrix}
     +&(+,1)&+\\
     -&+&-\\
     -&(-,3)&+
    \end{matrix}\right]\!,\;\;
\state 3\!:\! \left[\begin{matrix}
     +&-&-\\
     (+,1)&+&+\\
     (-,2)&-&+
    \end{matrix}\right]\!.
\end{equation}
Those tables define the model's behavior in the following way:
If e.g.\ the system is in state $\state 1$ and we measure the observable 
 $\gamma$, consider the first table at the position of $\gamma$ (i.e., the last 
 entry in the third row).
The $+$ sign at this position indicates that the measurement result will be 
 $+1$, while the system stays in state $\state 1$.
If we continue and measure $C$, we encounter the entry $(+, 2)$ which indicates 
 the measurement result $+1$ and a subsequent change to the state $\state 2$.
Being in state $\state 2$, the second table defines the behavior for the next 
 measurement: For instance a measurement of $c$ yields now the result $-1$ and 
 the system stays in state $\state 2$.

Thus, starting in state $\state 1$, the measurement results for the sequence 
 $\gamma C c$ are $+1, +1, -1$, so that the product is $-1$ in accordance with 
 the quantum prediction.
It is straightforward to verify that this model yields $\eval\chi= 6$.
In addition, the observables within each context are compatible in the sense 
 that in sequences of the form $AA$, $ABA$, or $A\alpha aA$, the first and last 
 measurement of $A$ yields the same output.
In fact this particular model is \emph{memory-optimal} (cf.\ 
 Theorem~\ref{t25868}) and we assign the symbol $\taurus$ to it.

\section{Memory cost of classical models}
Any model that reproduces contextuality eventually predicts that the system 
 attains different states during some measurement sequences.
As an omniscient observer one would know the state prior to each measurement 
 and could include it in the measurement record.
Thus, knowing the state of the system, one can predict the measurement outcome 
 as well as the state of the system that will occur prior to the next 
 measurement.
Thus we can write any model that explains the outcomes of sequential 
 measurements in the same fashion as we did for $\taurus$.

Taking a different point of view, any such model can be considered to be an 
 automaton with finitely many internal states, taking inputs (measurement 
 settings) and yielding outputs (measurement results).
In our notation, the output depends not only on the internal state, but also on 
 the input.
Such automatons are known as Mealy machines \cite{Mealy:1955BST, RothJr:2009}.

The quantum predictions add restrictions on such an automaton and thus increase 
 the number of internal states needed.
As a simple example we could require that an automaton reproduces the quantum 
 predictions from the rows and columns in the PM square.
That is, for all sequences in the set
\begin{equation}\label{e16648}
 \sqset_\mathrm{rc}= \{ABC, abc, \alpha\beta\gamma, Aa\alpha, Bb\gamma, 
 Cc\gamma, \text{  and permutations}\},
\end{equation}
 we require that the automaton must yield an output that matches the quantum 
  prediction.
For example, QM predicts for the sequence $ABC$ that the output is either $+1, 
 +1, +1$ or one of the permutations of $+1, -1, -1$.

More generally, if $\sqset$ denotes a set of measurement sequences, we say that 
 an automaton $\automaton$ \emph{obeys} the set $\sqset$ if the output for any 
 sequence in $\sqset$ matches the quantum prediction --- i.e., if for any 
 sequence in $\sqset$, the output of $\automaton$ could have occurred with a 
 nonvanishing probability according to the quantum scenario.
We say that a sequence \emph{yields a contradiction} if the output of this 
 sequence cannot occur according to QM.
Hereby we consider all quantum predictions from any initial state (it would 
 also suffice to only consider the completely mixed state $\varrho= 
 \openone/\tr[\openone]$).
Furthermore, we assume that prior to the measurement of a sequence, the 
 automaton always is re-initialized.
This ensures that the output of the automaton is independent of any action 
 prior to the selected measurement sequence.
Note that we only consider the \emph{certain} quantum predictions which occur 
 with a probability of {1}, while e.g.\ in the PM square we do not require that 
 for the sequence $A\gamma$ the probability of obtaining $+1, +1$ is equal to 
 the probability of obtaining $+1, -1$.
Finally, if an automaton with $k$ states $\state 1, \dotsc, \state k$ obeys 
 $\sqset$ and there exists no automaton with less states obeying $\sqset$, we 
 define the \emph{memory cost} of $\sqset$ to be $\mcost(\sqset)= \log_2(k)$.

\section{Contextuality conditions}
Our definition of memory cost so far applies to arbitrary situations, even 
 those in which contextuality does not directly play a role.
In contrast, contextuality of sequential measurements corresponds to the 
 particular feature, that certain sequences of mutually compatible observables 
 cannot be explained by a model with preassigned values (cf.\ 
 Ref.~\cite{Guhne:2010PRA} for a detailed discussion).
The contextuality conditions for observables $X_1, X_2, \dotsc$ thus arise from 
 the set of all sequences of mutually compatible observables,
\begin{equation}\label{e5303}
 \sqset_\mathrm{context}= \{ X_1 X_2 \dotsc \mid \text{$X_\ell$ mutually
                             compatible} \}.
\end{equation}
If the choice of observables $X_1, X_2, \dotsc$ exhibits contextuality, then 
 $\mcost(\sqset_\mathrm{context})>0$.
---
In the case of the PM square, $\sqset_\mathrm{context}$ surely contains all the 
 row and column sequences that we included in $\sqset_\mathrm{rc}$.
In addition, however, $\sqset_\mathrm{context}$ contains e.g.\ the sequences 
 $AA$, $ABA$, and $A\alpha aA$, for which QM predicts with certainty a 
 repetition of the value of $A$ in the first and last instance.
Note, that the set $\sqset_\mathrm{context}$ also contains more complicated 
 sequences like $ACABCA$ for which QM predicts with certainty that the values 
 of $A$ ($C$) in the first, third and sixth (second and fifth) measurement 
 coincide, and that product of the outcome for $ABC$ yields $+1$.

A particular feature of contextuality is that one can find observables that 
 exhibit contextuality independent of the actual preparation (the initial 
 state) of the quantum system.
Consequently, one may consider an extended preparation procedure of the 
 automaton, where the experimenter performs additional measurements between the 
 initialization of the automaton and the actual sequence.
The experimenter would e.g.\ measure the sequence $bABC$ but consider the 
 measurement of the observable $b$ to be actually part of the preparation 
 procedure.
We write $\ignore{b} ABC$ for a sequence where we are not interested in the 
 result of $b$.
If $\sqset_\mathrm{all}$ denotes the set of all sequences with observables 
 $X_1, X_2, \dotsc$, we write
\begin{equation}
 \sqset'_\mathrm{context}= \{ \ignore{T}S \mid S\in \sqset_\mathrm{context}, 
 T\in \sqset_\mathrm{all} \}
\end{equation}
 for the set of all sequences in $\sqset_\mathrm{context}$, including arbitrary 
 preparation procedures.

For the contextuality in the PM square, the automaton $\taurus$ obeys 
 $\sqset_\mathrm{context}$, while no automaton with less than {3} states can 
 obey $\sqset_\mathrm{context}$, cf.\ Appendix~\ref{X:aptaurusok} for details.
We did not specify an initial state for $\taurus$ and indeed the contextuality 
 conditions are obeyed for any initial state.
We summarize:
\begin{theorem}\label{t25868}
The memory cost for the contextuality correlations $\sqset'_\mathrm{context}$ 
 in the PM square is $\log_2(3)\approx 1.58$ bits; 
 $\mcost(\sqset'_\mathrm{context})= \mcost(\sqset_\mathrm{context})= 
 \log_2(3)$.
Consequently, the automaton $\taurus$ is memory-optimal.
\end{theorem}

\section{Compatibility conditions}
The set $\sqset_\mathrm{context}$ contains all sequences of mutually compatible 
 observables, but does not contain sequences like $ABaA$, for which QM also 
 predicts that both occurring values of $A$ are the same.
Sequences of this form enforce that all observables compatible with an 
 observable $Y$ must not change the measurement result of $Y$.
This can be covered by the set of all compatibility conditions
\begin{equation}\label{e27237}
 \sqset_\mathrm{compat}= \{ Y\ignore{X_1X_2\dotsc}Y \mid X_\ell
 \text{ compatible to } Y \},
\end{equation}
 and a convincing test of contextuality must also test the correlations due to 
 this set of sequences.
Again we define $\sqset'_\mathrm{compat}$ to include arbitrary preparation 
 procedures.

The automaton $\taurus$ does not obey $\sqset_\mathrm{compat}$, since e.g.\ 
 starting with state $\state 1$, the sequence $B\ignore{C\beta} B$ yields the 
 record $+1, \ignore{+1, -1,} {-1}$ and hence violates the assumption of 
 compatibility; similar sequences can be found for any initial state.
We show in Appendix~\ref{X:apnothree} that no automaton with three states can 
 obey simultaneously $\sqset'_\mathrm{compat}$ and $\sqset'_\mathrm{context}$ 
 and hence $\mcost(\sqset'_\mathrm{compat}$ and $\sqset'_\mathrm{context}) \ge 
 2$.
However, there exist automata with four internal states which obey 
 $\sqset'_\mathrm{compat}$ and $\sqset'_\mathrm{context}$.
As an example of such an automaton we define $\gemini$ via
\begin{equation}
\begin{split}
 \state 1\!:\! \left[\begin{matrix}
  +&+&(+,2)\\
  +&+&(+,3)\\
  \phantom{1}+\phantom{1}&\phantom{1}+\phantom{1}&+
 \end{matrix}\right]\!,
 & \;\;
 \state 2\!:\! \left[\begin{matrix}
  +&+&+\\
  -&+&-\\
  (-,4)&(+,1)&\phantom{1}+\phantom{1}
 \end{matrix}\right]\!,
 \\
 \state 3\!:\! \left[\begin{matrix}
  +&-&-\\
  +&+&+\\
  (+,1)&(-,4)&\phantom{1}+\phantom{1}
 \end{matrix}\right]\!,
 &\;\;
 \state 4\!:\! \left[\begin{matrix}
  +&-&(-,3)\\
  -&+&(-,2)\\
  \phantom{1}-\phantom{1}&\phantom{1}-\phantom{1}&+
 \end{matrix}\right].
\end{split}\end{equation}
Similar to the situation for $\taurus$, the initial state for the automaton 
 $\gemini$ can be chosen freely; we refer to Appendix~\ref{X:apgeminiok} for 
 details.
So we have:
\begin{theorem}
The memory cost of for the contextuality and compatibility correlations in the 
 PM square is two bits; $\mcost(\sqset'_\mathrm{compat} \text{ and } 
 \sqset'_\mathrm{context})= 2$.
Consequently, the automaton $\gemini$ is memory-optimal.
\end{theorem}

\section{Extended Peres-Mermin square}
There are, however, further contextuality effects for two qubits which then 
 require more than two bits for a simulation.
Namely, in Ref.~\cite{Cabello:2010PRA} an extension of the PM square has been 
 introduced, involving {15} different observables in {15} different contexts.
The argument goes as follows:
Consider the {15} observables of the type $\sigma_\mu \otimes \sigma_\nu$ where 
 $\mu, \nu \in \{ 0,x,y,z \}$ and $\sigma_0=\openone$ and the case $\mu=\nu=0$ 
 is excluded.
In this set there are {12} trios of mutually compatible observables such that 
 the product of their results is always $+1$, like $[\sigma_x \otimes \openone, 
 \openone \otimes \sigma_y, \sigma_x \otimes\sigma_y]$ and $[\sigma_x 
 \otimes\sigma_y, \sigma_y \otimes\sigma_x, \sigma_z \otimes\sigma_z]$, and 
 three trios of mutually compatible observables such that the product of their 
 results is always $-1$, like $[\sigma_x \otimes\sigma_y, \sigma_y 
 \otimes\sigma_z, \sigma_z \otimes\sigma_x]$.
This leads to {15} contexts in total.
Similar to the usual PM square, one can derive an state independent inequality.
For this inequality, QM predicts a value of {15} for the total sum, while 
 noncontextual models have a maximal value of {9}, cf.\ 
 Ref.~\cite{Cabello:2010PRA} and Appendix~\ref{X:apnofour} for details.

One may argue that this new contextuality argument is stronger than the usual 
 PM square \cite{Cabello:2010PRA}.
Does a simulation of it require more memory than the original PM square?
Indeed, this is the case:
\begin{theorem}
The memory cost for the contextuality and compatibility correlations in the 
 extended version of the PM square is strictly larger than two bits.
\end{theorem}

More precisely, according to Eqns.~\eqref{e5303} and \eqref{e27237} we define 
 the contextuality sequences $\sqset'_\mathrm{compat, 15}$ and compatibility 
 sequences $\sqset'_\mathrm{context, 15}$ for the {15} observables in the 
 extension of the PM square.
Then, the theorem states that $\mcost(\sqset'_\mathrm{compat, 15}$  and 
 $\sqset'_\mathrm{context, 15})>2$.
The proof is based on the following idea:
If one considers the {15} contexts in the extended square then they can be 
 arranged in a collection of {10} distinct squares, each similar to the usual 
 PM square.
The contextuality in this arrangement is strong enough such that for each fixed 
 assignment of the output, one must have \emph{three} contradictions for one of 
 the {10} usual PM squares.
One can show, however, that any four-state solution obeying 
 $\sqset'_\mathrm{context}$ and $\sqset'_\mathrm{compat}$ is similar to 
 $\gemini$, in which for no fixed state one has three contradictions.
The full proof is given in Appendix~\ref{X:apnofour}.

Although in this paper we are mainly concerned about the memory cost of 
 contextuality, we mention that the simulation of all certain quantum 
 predictions of the PM square already requires more than two bits of memory.
In fact, any four-state automaton that obeys $\sqset'_\mathrm{compat}$ and 
 $\sqset'_\mathrm{context}$ is up to some symmetries of the form of $\gemini$ 
 (cf.\ Appendix~\ref{X:apnofour}, Proposition~\ref{p7285}), but for $\gemini$ 
 the sequence $\ignore\beta ab\ignore Cc$ yields a contradiction.
This proves that $\mcost(\sqset_\mathrm{all})> 2$.

On the other hand, QM itself suggests an automaton for simulating 
 contextuality.
If e.g.\ we choose the pure state $\proj \phi$ defined by $A\ket\phi= B 
 \ket\phi = \ket\phi$ as initial state, then this state and all the states 
 occurring during measurement sequences define an (nondeterministic) automaton.
By a straightforward calculation one finds that this automaton attains {24} 
 different states if we consider the set of all sequences 
 $\sqset_\mathrm{all}$.
By a suitable elimination of the nondeterminism, we can readily reduce the 
 number of states to {10} (cf.\ Appendix~\ref{X:apten}) yielding an upper bound 
 on the required memory and hence $2< \mcost(\sqset_\mathrm{all})\le 
 \log_2(10)\approx 3.32$.

\section{Conclusions}
We have investigated the amount of memory needed in order to simulate quantum 
 contextuality in sequential measurements.
We determined the memory-optimal automata for important cases and have proven 
 that the simulation of contextuality phenomena for two qubits requires more 
 than two classical bits of memory.
However, the maximal amount of classical information that can be stored and 
 retrieved in two qubits is well known to be limited to two bits 
 \cite{Holevo:1973PIT}.
This implies that any classical model of such a system either would allow to 
 store and retrieve more than two bits or would have inaccessible degrees of 
 freedom.
(An example of the latter is $\taurus$, since one cannot perfectly infer the 
 initial state from the results of any measurement sequence.)

It should be emphasized that our analysis is about the memory that is needed to 
 classically simulate the \emph{certain} predictions from measurement 
 \emph{sequences} on a quantum system.
In contrast, one may ask how many different states are needed to merely explain 
 the observed expectation values \cite{Harrigan:2007XXX, Dakic:2008PRL, 
 Vertesi:2009PRA, Brukner:2009FP, Galvao:2009PRA, Gallego:2010PRL}.
However, the number of states needed in this scenario measures the number of 
 different initial configurations of the system, while we have shown that even 
 for a fixed initial configuration, the system must eventually attain a certain 
 number of states during measurement sequences.
Similarly, it has been demonstrated that a hybrid system of one qubit and one 
 classical bit of memory is \emph{on average} superior to a classical system 
 having only access to a single bit of memory \cite{Brukner:2004XXX} -- while 
 we show in Theorem~\ref{t25868}, that for a two-qubit system, even the certain 
 predictions cannot be simulated with one classical bit of memory.

Our work provides a link between information theoretical concepts on the one 
 side and quantum contextuality and the Kochen-Specker theorem on the other 
 side.
While for Bell's theorem such connections are well explored and have given deep 
 insights into QM \cite{Buhrman:2010RMP, Brassard:2006PRL, Pawlowski:2009NAT}, 
 for contextuality many questions remain open:
If an experiment violates some noncontextuality inequality up to a certain 
 degree, but not maximally, what memory is required to simulate this behavior?
Can nondeterministic machines help to simulate contextuality?
What amount of memory and randomness is required to simulate all quantum 
 effects in the PM square, especially in the distributed setting 
 \cite{Horodecki:2010XXX}?
Finally, for quantum non-locality it has been extensively investigated why QM 
 does not exhibit the maximal non-locality \cite{Brassard:2006PRL, 
 Pawlowski:2009NAT}.
A similar situation occurs for quantum contextuality --- can concepts from 
 information theory also help to understand the nonmaximal violation in this 
 situation?

\section*{Acknowledgments}
The authors thank
E.\ Amselem,
P.\ Badzi\c{a}g,
J.\ Barrett,
I.\ Bengtsson,
M.\ Bourennane,
\v{C}.\ Brukner,
P.\ Horodecki,
A.\ R.\ Plastino,
M.\ R{\aa}dmark, and
V.\ Scholz for discussions.
This work has been supported by the Austrian Science Fund (FWF): Y376 N16 
 (START Prize) and SFB FOQUS, the MICINN Projects MTM2008-05866 and 
 FIS2008-05596, the Wenner-Gren Foundation, and the EU (QICS, NAMEQUAM, Marie 
 Curie CIG 293993/ENFOQI).

\appendix
\section{$\taurus$ is optimal}\label{X:aptaurusok}
We already defined the set of row and column sequences $\sqset_\mathrm{rc}$ in 
 Eq.~\eqref{e16648}.
Another natural constraint is given by the set of repeated measurements
\begin{equation}
 \sqset_\mathrm{repeat}=
   \{AA, BB, CC, aa, bb, cc, \alpha\alpha, \beta\beta, \gamma\gamma\},
\end{equation}
 where we expect for any of these pairs that the results in the first and the 
 second measurement coincide.
Both sets $\sqset_\mathrm{rc}$ and $\sqset_\mathrm{repeat}$ obviously are 
 subsets of the set of contextuality sequences $\sqset_\mathrm{context}$ of the 
 PM square.
Nevertheless, an automaton that simultaneously obeys $\sqset_\mathrm{rc}$ and 
 $\sqset_\mathrm{repeat}$ already possesses more than two internal states, 
 i.e.\ $\mcost(\sqset_\mathrm{rc} \text{ and } \sqset_\mathrm{repeat})> 1$.
In order to see this, assume that the automaton has only two internal states 
 and without loss of generality that it starts in state $\state 1$.
We consider the case where in the last column there must be a prescribed state 
 change in order to avoid a contradiction, i.e., in $\state 1$ the product of 
 the assignments of $Cc\gamma$ is $+1$, contrary to the quantum prediction.
Note that there always exists at least one row or column with such a 
 contradiction, and that the proof for any row or column follows the same 
 lines.
If there is only one state change (say after a measurement of $\gamma$), then, 
while measuring the sequence $Cc\gamma$, the automaton would remain in $\state 
1$ until after the last output and therefore yield a contradiction.
If there are two (or more) state changes in the last column (say $c$ and 
 $\gamma$), both must go to $\state 2$.
Then, the constraints from $\sqset_\mathrm{repeat}$ require that $\gamma$ has 
 the same values in $\state 1$ and $\state 2$ (this is also true for $c{}$).
But then the sequence $C c \gamma$ in $\sqset_\mathrm{rc}$ will yield a 
 contradiction.
Thus a two-state automaton cannot obey both, $\sqset_\mathrm{rc}$ and 
 $\sqset_\mathrm{repeat}$.

On the other hand, $\taurus$ is an example of a {3} state automaton, which 
 obeys $\sqset_\mathrm{rc}$ and $\sqset_\mathrm{repeat}$.
In fact $\taurus$ obeys $\sqset'_\mathrm{context}$.
In order to see this, it is enough to show, that for any choice of the initial 
 state, the automaton will obey $\sqset_\mathrm{context}$.
So we assume that $\state 1$ is the initial state; the reasoning for $\state 2$ 
 and $\state 3$ is similar.
If we now measure a sequence with observables from the first row only, we may 
 jump between the states $\state 1$ and $\state 2$, but the output for all 
 observables in the first row are the same for either state.
A similar argument holds for all rows and the first and second column.
For a sequence with measurements from the third column, assume that the first 
 observable in the sequence, that is not $\gamma$, is the observable $c$.
Then the state changes to $\state 3$, in which the last column does not yield a 
 contradiction.
Since only the output $C$ was changed, but $C$ was not measured so far, we 
 cannot get any contradiction.
A similar argument can be used for the case where the first observable in the 
 sequence, that is not $\gamma$, is the observable $C$.

In summary, since any automaton that obeys $\sqset_\mathrm{context}$ has at 
 least {3} states and $\taurus$ is a {3} state automaton obeying the larger set 
 $\sqset_\mathrm{context}'$, we have shown that $\taurus$ is memory-optimal for 
 either set.

\section{$\gemini$ obeys $\sqset'_\mathrm{context}$ and 
          $\sqset'_\mathrm{compat}$}\label{X:apgeminiok}
In this Appendix we demonstrate that the automaton $\gemini$ indeed obeys 
 $\sqset'_\mathrm{context}$ and $\sqset'_\mathrm{compat}$.
The proof for $\sqset'_\mathrm{context}$ is completely analogous to the one in 
 Appendix~\ref{X:aptaurusok}.

For $\sqset'_\mathrm{compat}$, we consider a fixed observable, e.g.\ $B$.
Then $\state 1$ and $\state 2$ yield $+1$ while $\state 3$ and $\state 4$ give 
 $-1$.
However, using arbitrary measurements compatible to $B$, (i.e., $A$, $B$, $C$, 
 $b$, and $\beta$) we can never reach $\state 3$ or $\state 4$ if we start from 
 $\state 1$ or $\state 2$ and \emph{vice versa}.
Hence no contradiction occurs for any sequence of the type $\ignore T B 
 \ignore{X_1X_2\dotsc} B$.
A similar argument holds for all observables, if we note in addition that e.g.\ 
 after a measurement of $C$ the automaton can only be in $\state 2$ or $\state 
 3$.

\section{Definitions and basic rules used in the optimality 
proofs}\label{X:apdefinitions}
As we already did in the main text, we denote the observables from the PM 
 square by
\begin{equation}
\begin{bmatrix}
A & B & C \\
a & b & c \\
\alpha & \beta & \gamma
\end{bmatrix}.
\end{equation}
Furthermore, we denote the rows of the square by $\RR_i$, and the columns by 
 $\CC_i$.
The value table of each memory state $i$ is denoted by $T_i$ and the update 
 table by $U_i$.
We write an entry of zero in $U_i$, if the state does not change for that 
 observable.
Furthermore, we write measurement sequences as $A_1^+ B_2^- C_2^- a_3^+$ 
 meaning that when the sequence $ABCa$ was measured, the results were 
 $+,-,-,+$, and the memory was initially in state $\state 1$ and changed like 
 $\state 1 \mapsto \state 2 \mapsto \state 2 \mapsto \state 3$.

It will be useful for our later discussion to note some rules about the 
 structure of the value and update tables.
\begin{enumerate}
\item[1.] \emph{Sign flips:}
Let us assume that we have an automaton obeying $\sqset'_\mathrm{context}$ and 
 $\sqset'_\mathrm{compat}$ (or some subset of those sets) and pick a $2 \times 
 2$ square of observables (e.g.,~the set $\{A,B,a,b\}$ or 
 $\{A,B,\alpha,\beta\}$ or $\{A,C,\alpha,\gamma\}$).
Then, if we flip in each $T_i$ the signs corresponding to these observables, we 
 will obtain another valid automaton.

This holds true, because the mentioned sign flips do not change any of the 
 certain quantum predictions from $\sqset'_\mathrm{context}$ or
 $\sqset'_\mathrm{compat}$.
This rule will allow us later to fix one or two entries in a given value table 
 $T_i$.
\item[2.] \emph{Number of contradictions:}
Any table $T_i$ contains either one, three, or five contradictions to the row 
 and column constraints.

This follows directly from that fact that any fixed assignment fulfills 
 $\prod_k \RR_k \CC_k = +1$, while the row and column constraints require 
 $\prod_k \RR_k \CC_k = -1$.
\item[3.] \emph{Condition for fixing the memory:}
Let us assume that we have an automaton obeying $\sqset'_\mathrm{context}$ and 
 let there be a table $T_i$ which assigns to an observable (say $A$) a value 
 different from all other tables.
Then, the update table $U_i$ must contain only zeroes in the corresponding row 
 and column (here, $\RR_1$ and $\CC_1$).

The observables in the row and column correspond to compatible observables, 
 which are not allowed to change the value of the first observable.
However, any change of the memory state would change the value, as $T_i$ is the 
 only table with the initial assignment.
\item[4.] \emph{Contradictions and transformations:}
Let us assume that we have an automaton obeying $\sqset'_\mathrm{context}$ and 
 let there be in $T_i$ some contradiction in the column $\CC_j$ (or the row 
 $\RR_j$).
Then, in the update table $U_i$ there cannot be two zeroes in the the column 
 $\CC_j$ (or the row $\RR_j$).

If there were two zeroes, it could happen that one measures two entries of 
 $\CC_j$ without changing the memory state.
But then measuring the third one will reveal the contradiction in $T_i$.
(Note that the automaton first provides the result and then updates its state.)
\item[5.] \emph{Contradictions and other tables:}
Let us assume that we have an automaton obeying $\sqset'_\mathrm{context}$ and 
 let there be in $T_i$ a contradiction in the column $\CC_j$ (or the row 
 $\RR_j$).
Then, there must be two different tables $T_{k}$ and $T_l$ where in both the 
 column $\CC_j$ has no contradictions anymore, but the assignments of $T_{k}$ 
 and $T_l$ differ in two observables of $\CC_j$.
Furthermore, in the column $\CC_j$ of the update table $U_i$ there must be two 
 entries leading to two different states.

First, note that there must be at least one other table $T_{k}$ where the 
 contradiction does not exist anymore.
This follows from the fact that we may measure $\CC_j$ starting from the memory 
 state $i$.
After having made these measurements, we arrive at some state $k$, and from the 
 contextuality correlations $\sqset'_\mathrm{context}$ it follows that $\CC_j$ 
 in $T_{k}$ has no contradiction.

The table $T_{k}$ differs from $T_i$ in at least one observable $X$ in $\CC_j$.
On the other hand, starting from $T_i$ one might measure $X$ as a first 
 observable.
Then, making further measurements on $\CC_j$ one must arrive at a table $T_l$ 
 without a contradiction.
Since $T_{k}$ and $T_l$ have both no contradiction, they must differ in at 
 least two places, one of them being $X$.
Finally, if the column $\CC_j$ in $U_i$ would only have entries of zero and 
 $k$, then $\CC_j$ in $T_k$ could not differ from $T_i$.
This eventually leads to a contradiction and hence proves the last assertion.
\end{enumerate}

\section{$\gemini$ is memory-optimal}\label{X:apnothree}
Here, we proof the optimality of the 4-state automaton $\gemini$, in the sense 
 of obeying $\sqset'_\mathrm{context}$ and $\sqset'_\mathrm{compat}$ with a 
 minimum number of states.
We use the definitions and rules as introduced in 
Appendix~\ref{X:apdefinitions}.

Let us assume that we would have a three-state automaton obeying 
 $\sqset'_\mathrm{context}$.
$T_1$ has a contradiction, and we can assume, without loss of generality, that 
 it is $\CC_3$.
Then, according to Rule~1 we can, without loss of generality, assume that all 
 entries in $\CC_3$ are ``+''.
Together with Rule~5 this leads to the conclusion that the three states $T_i$ 
 are, without loss of generality, of the form:
\begin{subequations}
\begin{align}\label{threestates}
T_1&\!:\! \begin{bmatrix}
 \phantom{-} & \phantom{-} & + \\
             &             & + \\
             &             & +
\end{bmatrix}\!,&\!\!\!\!
T_2&\!:\! \begin{bmatrix}
 \phantom{+} & \phantom{-} & + \\
             &             & + \\
             &             & -
\end{bmatrix}\!,&\!\!\!\!
T_3&\!:\! \begin{bmatrix}
 \phantom{+} & \phantom{-} & + \\
             &             & - \\
             &             & +
\end{bmatrix}\!,\\
U_1&\!:\! \begin{bmatrix}
 \phantom{0} & \phantom{0} & \phantom{0}  \\
             &             & 2            \\
             &             & 3
\end{bmatrix}\!,&\!\!\!\!
U_2&\!:\! \begin{bmatrix}
 \phantom{0} & \phantom{0} &  0 \\
             &             &  0 \\
           0 & 0           &  0
\end{bmatrix}\!,&\!\!\!\!
U_3&\!:\! \begin{bmatrix}
 \phantom{0} & \phantom{0} & 0 \\
           0 & 0           & 0 \\
             &             & 0
\end{bmatrix}\!.
\end{align}
\end{subequations}
Here, empty places in the tables mean that the corresponding entries are not 
 yet fixed.
The table $U_1$ follows from Rule~5, and $U_2$ and $U_3$ follow from Rule~3.

Which can be the entries corresponding to the observables $a$ and $b$ in $U_2$?
Since $T_3$ assigns a different value to $c$ than $T_2$, there cannot be a 
 ``3'' at these entries, otherwise, a sequence like $c_2^+ a_2^? c_3^-$ would 
 lead to a contradiction to the conditions of $\sqset_\mathrm{context}$.

But there can also not be a ``1'' at these entries, because then the sequence 
 $c_2^+ a_2^? \gamma_1^+ c_3^-$ yields a contradiction to 
 $\sqset_\mathrm{compat}$, since $a$ and $\gamma$ are compatible with $c$.
So the entries of $\RR_2$ in $U_2$ must be zero, as there are only three states 
 in the memory.
A similar argument can be applied to $U_3$, showing that here $\RR_3$ must be 
 zero.

So the tables have to be of the form
\begin{subequations}
\begin{align}
T_1&\!:\! \begin{bmatrix}
 \phantom{-} & \phantom{-} & + \\
             &             & + \\
             &             & +
\end{bmatrix}\!,&\!\!\!\!
T_2&\!:\! \begin{bmatrix}
 \phantom{+} & \phantom{-} & + \\
             &             & + \\
             &             & -
\end{bmatrix}\!,&\!\!\!\!
T_3&\!:\! \begin{bmatrix}
 \phantom{+} & \phantom{-} & + \\
             &             & - \\
             &             & +
\end{bmatrix}\!,\\
\label{zerotrick}
U_1&\!:\! \begin{bmatrix}
 \phantom{0} & \phantom{0} & \phantom{0} \\
             &             & 2 \\
             &             & 3
\end{bmatrix}\!,&\!\!\!\!
U_2&\!:\! \begin{bmatrix}
 \phantom{0} & \phantom{0} & 0 \\
           0 &           0 & 0 \\
           0 &           0 & 0
\end{bmatrix}\!,&\!\!\!\!
U_3&\!:\! \begin{bmatrix}
 \phantom{0} & \phantom{0} & 0 \\
           0 &           0 & 0 \\
           0 &           0 & 0
\end{bmatrix}\!.
\end{align}
\end{subequations}
Now, according to Rule~4, the contradictions in $T_2$ as well as in $T_3$ can 
 only be in $\RR_1$.
But, according to Rule~5, if there is a contradiction in $\RR_1$ of $T_2$, 
 there must be two different $T_i$ and $T_j$ where there is no contradiction in 
 $\RR_1$.
But there is only one table left, namely $T_1$, and we arrive at a 
 contradiction.

\section{Proof that the classical simulation of the extended PM square requires 
          more than two bits of memory}\label{X:apnofour}
Let us now discuss the extended PM square from Ref.~\cite{Cabello:2010PRA}.
Again, we refer to Appendix~\ref{X:apdefinitions} for basic definitions and 
rules.
As already mentioned,
one considers for that the array of observables
\begin{equation}
\begin{bmatrix}
& \chi_{01} & \chi_{02}& \chi_{03}\\
\chi_{10}& \chi_{11} & \chi_{12}& \chi_{13}\\
\chi_{20}& \chi_{21} & \chi_{22}& \chi_{23}\\
\chi_{30}& \chi_{31} & \chi_{32}& \chi_{33}
\end{bmatrix}
=
\begin{bmatrix}
& \openone \otimes \sigma_x & \openone \otimes \sigma_y& \openone \otimes \sigma_z
\\
\sigma_x \otimes \openone & \sigma_x \otimes\sigma_x & \sigma_x \otimes\sigma_y &\sigma_x \otimes\sigma_z
\\
\sigma_y \otimes \openone & \sigma_y \otimes\sigma_x & \sigma_y \otimes\sigma_y &\sigma_y \otimes\sigma_z
\\
\sigma_z \otimes \openone & \sigma_z \otimes\sigma_x & \sigma_z \otimes\sigma_y &\sigma_z \otimes\sigma_z
\end{bmatrix}\!.
\end{equation}
These observables can be grouped into trios, in which the observables commute 
and their product equals $\pm \openone.$ Nine trios are of the form $\{ 
\chi_{k0}, \chi_{kl}, \chi_{0l}\}$, three trios where the product equals 
$+\openone$ are $\{ \chi_{11}, \chi_{23},\chi_{32}\}$, $\{ \chi_{12}, 
\chi_{21},\chi_{33}\}$,
and
$\{ \chi_{13}, \chi_{22},\chi_{31}\}$.
Three trios where the product equals $-\openone$ are $\{ \chi_{11}, 
\chi_{22},\chi_{33}\}$, $\{ \chi_{12}, \chi_{23},\chi_{31}\}$,
and
$\{ \chi_{13}, \chi_{21},\chi_{32}\}$.
{From} this, one can derive the inequality
\begin{multline}
\textstyle\sum_{k,l}  \eval{\chi_{k0} \chi_{kl} \chi_{0l}}
+ \eval{ \chi_{11} \chi_{23}\chi_{32}} + \eval{ \chi_{12} \chi_{21}\chi_{33}}+
\\
 +\eval{\chi_{13} \chi_{22}\chi_{31}} - \eval{\chi_{11} \chi_{22}\chi_{33}}
-\eval{ \chi_{12} \chi_{23}\chi_{31}}-
\\
-\eval{ \chi_{13}\chi_{21}\chi_{32}} \leq 9
\end{multline}
for noncontextual models, while QM predicts a value of 15, 
independently of the state.

First note that in this new inequality 15 terms (or contexts) occur but any 
noncontextual model can fulfill the quantum prediction for only 12 of them at 
most, so three contradictions cannot be avoided.  One can directly check that 
in the whole construction of the inequality, 10 different PM squares occur.  
Nine of them are a simple rewriting of the usual
PM square, while the tenth comes from the observables $\chi_{kl}$
with $k,l \neq 0.$  Any of the 15 terms in the inequality contributes to four 
of these PM squares.

Any value table for the 15 observables leads to assignments to the 15 contexts, 
but it has at least three contradictions.  As any context contributes to four 
PM squares, this would lead to 12 contradictions in the 60 contexts of the 10 
PM squares, if we consider them separately. Since in a PM square the number of 
contradictions cannot be two (Rule 2), this means that one of the PM squares 
has to have three contradictions.

Let us now assume that we have a valid automaton for this extended PM square 
with four memory states. Of course, this would immediately give a valid 
four-state automaton of any of the 10 PM squares. For one of these PM squares, 
at least one table has to have {\it three} contradictions.  So it suffices to 
prove the following Lemma:
\begin{lemma}
There is no four-state automaton obeying $\sqset'_{compat}$ and 
$\sqset'_{context}$, where one table $T_i$ has three contradictions.
\end{lemma}

In course of proving this Lemma we will also prove the following:
\begin{proposition}\label{p7285}
The four-state automaton $\gemini$ is unique, up to some permutation or sign 
changes.
\end{proposition}

To prove the Lemma, we proceed in the following way: Without loss of 
generality, we can assume that the first three tables $T_i$ look like the $T_i$ 
in
Eq.~(\ref{threestates}). Then, we can add a fourth table $T_4.$ For the  
last column of this table, there are $2^3=8$ possible values. We will 
investigate all eight possibilities and show that either we arrive
directly at a contradiction, or that only an automaton similar to $\gemini$ is 
possible, in which any table has only one contradiction.  This proves the 
Lemma.

We will first deal with the four cases, where $T_4$ has also a contradiction
in $\CC_3.$ This will lead to Observation~\ref{o28709}, which will be useful in 
the following four cases.

\noindent
\\
{\bf Case 1:} {\it For $T_4$ one has $[C,c,\gamma]=[+++]$:\\}
In this case, a simple application of the previous rules
implies that several entries are fixed:
\begin{equation}
T:
\begin{bmatrix}
\phantom{-} & \phantom{-} & + \\
 &  & + \\
 &  & + 
\end{bmatrix}\!,
\;\;
\begin{bmatrix}
\phantom{+} & \phantom{-} & + \\
 &  & + \\
 &  & - 
\end{bmatrix}\!,
\;\;
\begin{bmatrix}
\phantom{+}& \phantom{-} & + \\
 & & - \\
 & & + 
\end{bmatrix}\!,
\;\;
\begin{bmatrix}
\phantom{+}& \phantom{-} & + \\
 & & + \\
 & & + 
\end{bmatrix}\!,
\end{equation}
\begin{equation}
U:
\begin{bmatrix}
\phantom{0} & \phantom{0} & \phantom{0}  \\
 &  & 2 \\
 &  & 3 
\end{bmatrix}\!,
\;\;\;\;
\begin{bmatrix}
\phantom{0} & \phantom{0} & {0}  \\
0 & 0 & 0 \\
0 & 0 &  0
\end{bmatrix}\!,
\;\;\;\;
\begin{bmatrix}
\phantom{0} & \phantom{0} & {0}  \\
0 & 0 & 0 \\
0 & 0 &  0
\end{bmatrix}\!,
\;\;\;\;
\begin{bmatrix}
\phantom{0} & \phantom{0} & \phantom{0}  \\
 &  & 2 \\
 &  & 3 
\end{bmatrix}\!.
\end{equation}
Here and in the following, we write the $T_i$ and $U_i$ just as 
a row for notational simplicity, starting from $T_1$ to $T_4.$
The entries in $U_1$ and $U_4$ are fixed from the following reasoning: Let us 
assume that one measures $c$ in $T_1$, then, since the values $C(T_i)$ are the 
same in all $T_i$, one has to change immediately to a table with no 
contradiction in $\CC_3$, and where the value of $c$ is still the same. The 
only possibility is $T_2.$ Furthermore, $\RR_2$ in $U_2$ and $\RR_3$ in $U_3$ 
must be zero due to the same argument which led to Eq.~(\ref{zerotrick}).

It follows (Rule 4) that $T_2$ and $T_3$ have both exactly one 
contradiction, which must be in $\RR_1.$ So, in $\RR_1$ of $U_2$ 
there must be the entries ``1'' and ``4''  [an entry ``3'' would 
not solve the problem, because in $\RR_1(T_3)$ has also a contradiction].  As 
we can still permute the first and second column, we can without loss of 
generality assume that the first row in $U_2$ is $[1\;\; 4\;\; 0].$ Due to Rule 
1, we can also assume, without loss of generality, that $A(T_2)=+.$ Similarly, 
in $\RR_1(U_3)$ there must be the entries ``1'' and ``4'',
resulting in two different cases: 

If $\RR_1(U_3) = [1\;4\; 0],$ we must have the following 
tables,
\begin{equation}
T:
\begin{bmatrix}
{+} & \phantom{-} & + \\
 &  & + \\
 &  & + 
\end{bmatrix}\!,
\;\;
\begin{bmatrix}
{+} & {-} & + \\
 &  & + \\
 &  & - 
\end{bmatrix}\!,
\;\;
\begin{bmatrix}
{+}& {-} & + \\
 & & - \\
 & & + 
\end{bmatrix}\!,
\;\;
\begin{bmatrix}
\phantom{+}& {-} & + \\
 & & + \\
 & & + 
\end{bmatrix}\!,
\end{equation}
\begin{equation}
U:
\begin{bmatrix}
\phantom{0} & \phantom{0} & \phantom{0}  \\
 &  & 2 \\
 &  & 3 
\end{bmatrix}\!,
\;\;\;\;
\begin{bmatrix}
1 & 4 & {0}  \\
0 & 0 & 0 \\
0 & 0 &  0
\end{bmatrix}\!,
\;\;\;\;
\begin{bmatrix}
1 & 4 & {0}  \\
0 & 0 & 0 \\
0 & 0 &  0
\end{bmatrix}\!,
\;\;\;\;
\begin{bmatrix}
\phantom{0} & \phantom{0} & \phantom{0}  \\
 &  & 2 \\
 &  & 3 
\end{bmatrix}\!.
\end{equation}
where the added values in $\RR_1$ of the $T_i$ follow from 
$\RR_1(U_2)$  and $\RR_1(U_3).$

Now, if we start from $T_2$ and measure the sequence $a_2 A_2 a_1$,
we see that we must have $a(T_1)=a(T_2)$. Similarly, from $T_3$ we 
can measure $a_3 A_3 a_1$, implying that
$a(T_1)=a(T_2)= a(T_3)$. Similarly, we find that $b(T_2)=b(T_3)= b(T_4)$. 
But this gives a contradiction: In $\RR_2(T_2)$ and $\RR_2(T_3)$ there is no 
contradiction and $c(T_2)\neq c(T_3)$. Therefore, it cannot be that $a(T_2)= 
a(T_3)$ {and} at the same time $b(T_2)=b(T_3)$.

As the second case, we have to consider the possibility that 
$\RR_1(U_3)=[4\;1\; 0]$. Then, also the values of $\RR_1(T_3)$
must be interchanged, $\RR_1(T_3) = [- +  +].$
Then, starting from $T_2$, the sequence $\alpha_2 A_2 \gamma_1 \alpha_3$ shows 
directly that $\alpha(T_2)=\alpha(T_3)$. Similarly, starting from $T_3$, the 
sequence $a_3 A_3 c_4 a_2$ shows that $a(T_2)=a(T_3).$
But since $A(T_2)\neq A(T_3)$, this is a contradiction.

\noindent
\\
{\bf Case 2:} {\it For $T_4$ one has $[C,c,\gamma]=[+--]$:\\}
As in Case 1, one can directly see that several entries are 
fixed:
\begin{equation}
T:
\begin{bmatrix}
\phantom{-} & \phantom{-} & + \\
 &  & + \\
 &  & + 
\end{bmatrix}\!,
\;\;
\begin{bmatrix}
\phantom{+} & \phantom{-} & + \\
 &  & + \\
 &  & - 
\end{bmatrix}\!,
\;\;
\begin{bmatrix}
\phantom{+}& \phantom{-} & + \\
 & & - \\
 & & + 
\end{bmatrix}\!,
\;\;
\begin{bmatrix}
\phantom{+}& \phantom{-} & + \\
 & & - \\
 & & - 
\end{bmatrix}\!,
\end{equation}
\begin{equation}
U:
\begin{bmatrix}
\phantom{0} & \phantom{0} & \phantom{0}  \\
 &  & 2 \\
 &  & 3 
\end{bmatrix}\!,
\;\;\;\;
\begin{bmatrix}
\phantom{0} & \phantom{0} & {0}  \\
0 & 0 & 0 \\
0 & 0 &  0
\end{bmatrix}\!,
\;\;\;\;
\begin{bmatrix}
\phantom{0} & \phantom{0} & {0}  \\
0 & 0 & 0 \\
0 & 0 &  0
\end{bmatrix}\!,
\;\;\;\;
\begin{bmatrix}
\phantom{0} & \phantom{0} & \phantom{0}  \\
 &  & 3 \\
 &  & 2 
\end{bmatrix}\!.
\end{equation}
The zeroes in $U_2$ and $U_3$ come from the following argumentation: 
Starting from $T_1$, the measurement sequence $c_1^+ X_2 c_?$ with 
$X$ compatible to $c$ shows that in $\RR_2(U_2)$ and $\CC_3(U_2)$
there can be no ``3'' or ``4''. But there can be also no ``1'', because 
then the  sequence $c_1^+ X_2 \gamma_1 c_3^-$ would lead to a contradiction.
Therefore, $\RR_2(U_2)$ and $\CC_3(U_2)$  have to be zero. Starting from $T_4$
and measuring $\gamma$ one can similarly prove that the entries for 
$\RR_3(U_2)$ have to be zero and analogous arguments prove also the zeroes in 
$U_3$.

It is now clear (Rule 4) that the contradictions in $T_2$ and $T_3$ have 
to be in $\RR_1$ and the missing entries in $U_2$ and $U_3$ can only be
``4'' and ``1''. As we still can permute the first and second column, 
there are only two possibilities:

{\bf Case 2A:}
First, we consider the case that $\RR_1(U_2)= \RR_1(U_3)= [1 \; 4 \; 0].$ 
As in Case 1, we can directly see that 
$a(T_2)=a(T_1)=a(T_3)$ and $b(T_2)=b(T_4)=b(T_3)$. Hence, 
$\RR_2(T_2)$ and $\RR_2(T_3)$ differ exactly in the value of $c$, but in both 
cases there is no contradiction in $\RR_2.$ This is not possible.

{\bf Case 2B:}
Second, we consider the case that the first rows of 
$U_2$ and $U_3$ differ, 
and we take $\RR_1(U_2)= [1 \; 4 \; 0]$ and $\RR_1(U_3)= [4 \; 1 \; 0].$
Then, we apply Rule 1 to fix for $A(T_3)=a(T_3)=+.$ 
Then, the tables have to be:
\begin{equation}
T:
\begin{bmatrix}
{-} & {-} & + \\
 & - & + \\
 & + & + 
\end{bmatrix}\!,
\;\;
\begin{bmatrix}
{-} & {+} & + \\
 &  & + \\
 &  & - 
\end{bmatrix}\!,
\;\;
\begin{bmatrix}
{+}& {-} & + \\
 +&- & - \\
 +&+ & + 
\end{bmatrix}\!,
\;\;
\begin{bmatrix}
{+}& {+} & + \\
 + & & - \\
 + & & - 
\end{bmatrix}\!,
\end{equation}
\begin{equation}
U:
\begin{bmatrix}
\phantom{0} & \phantom{0} & \phantom{0}  \\
 &  & 2 \\
 &  & 3 
\end{bmatrix}\!,
\;\;\;\;
\begin{bmatrix}
1 & 4 & {0}  \\
0 & 0 & 0 \\
0 & 0 &  0
\end{bmatrix}\!,
\;\;\;\;
\begin{bmatrix}
4 & 1 & {0}  \\
0 & 0 & 0 \\
0 & 0 &  0
\end{bmatrix}\!,
\;\;\;\;
\begin{bmatrix}
\phantom{0} & \phantom{0} & \phantom{0}  \\
 &  & 3 \\
 &  & 2 
\end{bmatrix}\!.
\end{equation}
Here, $\CC_2(T_1)$ and $\CC_1(T_4)$ come from measurement sequences 
like $a_3^+ A_3^+ a_4^+,$ starting from $T_3.$

Again, we have two possibilities for the value of 
$b$ in $T_2.$ If we set $b(T_2)=-$, then all values 
in all $T_i$ are fixed and each table has exactly one contradiction. This is, 
up to some relabeling,
the four-state automaton $\gemini$ from the main text (indeed, this  is the way 
how this solution was found). If we set $b(T_2)=+$, then also all $T_i$ can be 
filled, and we must have:
\begin{equation}
T:
\begin{bmatrix}
{-} & {-} & + \\
+ & - & + \\
- & + & + 
\end{bmatrix}\!,
\;\;
\begin{bmatrix}
{-} & {+} & + \\
 +&  +& + \\
 -& + & - 
\end{bmatrix}\!,
\;\;
\begin{bmatrix}
{+}& {-} & + \\
 +&- & - \\
 +&+ & + 
\end{bmatrix}\!,
\;\;
\begin{bmatrix}
{+}& {+} & + \\
 + &+ & - \\
 + &+ & - 
\end{bmatrix}\!,
\end{equation}
\begin{equation}
U:
\begin{bmatrix}
\phantom{0} & \phantom{0} & \phantom{0}  \\
 & 3 & 2 \\
2 &  & 3 
\end{bmatrix}\!,
\;\;\;\;
\begin{bmatrix}
1 & 4 & {0}  \\
0 & 0 & 0 \\
0 & 0 &  0
\end{bmatrix}\!,
\;\;\;\;
\begin{bmatrix}
4 & 1 & {0}  \\
0 & 0 & 0 \\
0 & 0 &  0
\end{bmatrix}\!,
\;\;\;\;
\begin{bmatrix}
\phantom{0} & \phantom{0} & \phantom{0}  \\
 & 2 & 3 \\
3 &  & 2 
\end{bmatrix}\!.
\end{equation}
Here, the tables $T_1$ and $T_4$ have three contradictions
(two new ones in $\RR_2$ and $\RR_3$) and the new entries 
in $U_1$ and $U_4$ must be introduced according to Rule 5
[note that $a(T_i)$ and $\beta(T_i)$ are for all tables the same].
Then, however,
starting from $T_1$, the sequence $\alpha_1^- A_2^- \gamma_1^+ \alpha_3^+$ 
shows that this is not valid solution.

\noindent
\\
{\bf Case 3:} {\it For $T_4$ one has $[C,c,\gamma]=[-+-]$:\\}
In this case, a simple reasoning according to the usual rules
fixes the entries:
\begin{equation}
T:
\begin{bmatrix}
\phantom{-} & \phantom{-} & + \\
 &  & + \\
 &  & + 
\end{bmatrix}\!,
\;\;
\begin{bmatrix}
\phantom{+} & \phantom{-} & + \\
 &  & + \\
 &  & - 
\end{bmatrix}\!,
\;\;
\begin{bmatrix}
\phantom{+}& \phantom{-} & + \\
 & & - \\
 & & + 
\end{bmatrix}\!,
\;\;
\begin{bmatrix}
\phantom{+}& \phantom{-} & - \\
 & & + \\
 & & - 
\end{bmatrix}\!,
\end{equation}
\begin{equation}
U:
\begin{bmatrix}
\phantom{0} & \phantom{0} & \phantom{0}  \\
 &  &  \\
 &  & 3 
\end{bmatrix}\!,
\;\;\;\;
\begin{bmatrix}
\phantom{0} & \phantom{0} & \phantom{0}  \\
 &  &  \\
 &  &  
\end{bmatrix}\!,
\;\;\;\;
\begin{bmatrix}
\phantom{0} & \phantom{0} & {0}  \\
0 & 0 & 0 \\
 &  &  0
\end{bmatrix}\!,
\;\;\;\;
\begin{bmatrix}
{0} & {0} & {0}  \\
 &  & 0 \\
 &  & 0 
\end{bmatrix}\!.
\end{equation}
Here we have an obvious contradiction in $T_4 / U_4$: $\CC_3(T_4)$ contains a 
contradiction, but (due to Rule 3) one is not allowed to change the memory 
state when measuring it. Therefore, the memory can never be in the state $4$.  
But then, one would have effectively a three-state solution, which is not 
possible, as we know already.

\noindent
\\
{\bf Case 4:} {\it For $T_4$ one has $[C,c,\gamma]=[--+]$:\\}
This is the same as Case 3, where $\RR_2$ and $\RR_3$ have been
interchanged.

\noindent
\\
Now we have dealt with all the cases, where $T_4$ contains a contradiction 
in $\CC_3,$ just as $T_1.$ We have seen that in this cases there can only 
be a solution if each table contains exactly one contradiction, and this solution
is unique, up to some permutations or sign flips. Moreover, we could have made 
the same discussion with rows instead of columns.
Therefore from the first four cases we can state an observation which will be 
useful in the remaining four cases:
\begin{observation}\label{o28709}
If in any four-state solution two tables $T_i$ and $T_j$ have both a 
 contradiction in the same column $\CC_k$ (or row $\RR_k$), then there has to 
 be exactly one contradiction in each value table of the automaton.
\end{observation}
So, if there is a four-state solution where
one table has three contradictions, then it cannot be that two tables have both 
a contradiction in the same column or row. 

Then we can proceed with the remaining cases.

\noindent
\\
{\bf Case 5:} {\it For $T_4$ one has $[C,c,\gamma]=[++-]$:\\}
This is the critical case, as it is difficult to distinguish the 
tables $T_2$ and $T_4$ here. First, the following entries 
are directly fixed:
\begin{equation}
T:
\begin{bmatrix}
\phantom{-} & \phantom{-} & + \\
 &  & + \\
 &  & + 
\end{bmatrix}\!,
\;\;
\begin{bmatrix}
\phantom{+} & \phantom{-} & + \\
 &  & + \\
 &  & - 
\end{bmatrix}\!,
\;\;
\begin{bmatrix}
\phantom{+}& \phantom{-} & + \\
 & & - \\
 & & + 
\end{bmatrix}\!,
\;\;
\begin{bmatrix}
\phantom{+}& \phantom{-} & + \\
 & & + \\
 & & - 
\end{bmatrix}\!,
\end{equation}
\begin{equation}
U:
\begin{bmatrix}
\phantom{0} & \phantom{0} & \phantom{0}  \\
 &  & 2 \\
 &  & 3 
\end{bmatrix}\!,
\;\;\;\;
\begin{bmatrix}
\phantom{0} & \phantom{0} & \scriptstyle{0|4}  \\
 \scriptstyle{0|4}& \scriptstyle{0|4} & \scriptstyle{0|4} \\
 &  & \scriptstyle{0|4}
\end{bmatrix}\!,
\;\;\;\;
\begin{bmatrix}
\phantom{0} & \phantom{0} & {0}  \\
0 & 0 & 0 \\
0 & 0 &  0
\end{bmatrix}\!,
\;\;\;\;
\begin{bmatrix}
\phantom{0} & \phantom{0} &   \\
 &  & \scriptstyle{0|2} \\
 &  & \scriptstyle{0|2}
\end{bmatrix}\!.
\end{equation}
Here,  $c(U_1)=2$ has been chosen without loss of generality. It is
clear that $c(U_1)=2$ or $c(U_1)=4$, as $T_2$ and $T_4$ are equivalent 
at the beginning, we can choose $T_2$ here. The entries of the type
$\scriptstyle{i|j}$  in $U_2$ and $U_4$ mean that the numbers can be $i$ or 
$j$, but nothing else. The values of $c(U_2)$
[and $c(U_4)$] cannot be 1, because then the sequence 
$c_2^+ \gamma_1^+ c_3^-$ directly reveals a contradiction.
Furthermore, the zeroes in $\RR_3(U_3)$ and 
$\RR_2(U_3)$ follow similarly as Eq.~(\ref{zerotrick}) or from Rule 3. 
In addition, $C(U_2)\neq 1$, because otherwise the sequence $c_1^+ C_2^+ 
\gamma_1^+ $ reveals a contradiction to the PM conditions.
Also, $C(U_2)\neq 3$, because of $c_1^+ C_2^+ c_3^-.$ Similarly, 1 and 3 are  
excluded as values for $a(U_2)$ and $b(U_2)$, due to the sequences $c_1^+ a_2 
\gamma_1^+ c_3^-$ and $c_1^+ a_2^+ c_3^-.$ 

Furthermore, we can use our Observation~\ref{o28709}: If in a four-state 
solution one column has a contradiction in two of the $T_i,$ then there can be 
only one contradiction in any $T_i$. Here we can use it as follows: It is clear 
that $T_3$ has its contradiction in $\RR_1$. Since we aim to rule out a 
four-state solution where one table has three contradictions,
we can assume that there is no contradiction in $\RR_1$ in all the
other $T_i$ (especially in $T_2$ and $T_3$). Otherwise, we would
already know that no solution exists with three contradictions
in a table.
We can distinguish two cases:

{\bf Case 5A:} Let us assume that $\gamma(U_2)=0.$ Then, the tables must read:
\begin{equation}
T:
\begin{bmatrix}
\phantom{-} & \phantom{-} & + \\
 &  & + \\
 &  & + 
\end{bmatrix}\!,
\;\;
\begin{bmatrix}
\phantom{+} & \phantom{-} & + \\
 &  & + \\
 &  & - 
\end{bmatrix}\!,
\;\;
\begin{bmatrix}
\phantom{+}& \phantom{-} & + \\
 & & - \\
 & & + 
\end{bmatrix}\!,
\;\;
\begin{bmatrix}
\phantom{+}& \phantom{-} & + \\
 & & + \\
 & & - 
\end{bmatrix}\!,
\end{equation}
\begin{equation}
U:
\begin{bmatrix}
\phantom{0} & \phantom{0} & \phantom{0}  \\
 &  & 2 \\
 &  & 3 
\end{bmatrix}\!,
\;\;\;\;
\begin{bmatrix}
\phantom{0} & \phantom{0} & \scriptstyle{0|4}  \\
\scriptstyle{0|4} & \scriptstyle{0|4} & \scriptstyle{0|4} \\
 \scriptstyle{0|4}& \scriptstyle{0|4} & 0
\end{bmatrix}\!,
\;\;\;\;
\begin{bmatrix}
\phantom{0} & \phantom{0} & {0}  \\
0 & 0 & 0 \\
0 & 0 &  0
\end{bmatrix}\!,
\;\;\;\;
\begin{bmatrix}
\phantom{0} & \phantom{0} &   \\
 &  & \scriptstyle{0|2} \\
 &  & \scriptstyle{0|2}
\end{bmatrix}\!.
\end{equation}
The new entries in $U_2$ follow from $\gamma(U_2)=0$ in combination with
$\gamma(T_1)=\gamma(T_3)\neq \gamma(T_2)$.

Due to Rule 5, the table $T_2$ must have a contradiction 
in $\CC_1$, $\CC_2$, or $\RR_1.$
{From} Observation~\ref{o28709}, we can assume that it is not in $\RR_1$.
Due to possible permutations of  $\CC_1$ and  $\CC_2$ we further
assume without loss of generality that the contradiction is in $\CC_1$.
Then we have:
\begin{equation}
U\!:\!
\begin{bmatrix}
\phantom{0} & \phantom{0} & \phantom{0}  \\
 &  & 2 \\
 &  & 3 \end{bmatrix}\!,\!\;
\begin{bmatrix}
1 & \phantom{0} & \scriptstyle{0|4}  \\
4 & \scriptstyle{0|4} & \scriptstyle{0|4} \\
 \scriptstyle{0|4}& \scriptstyle{0|4} & 0
\end{bmatrix}\!,
\;\;\;\;
\begin{bmatrix}
\phantom{0} & \phantom{0} & {0}  \\
0 & 0 & 0 \\
0 & 0 &  0
\end{bmatrix}\!,
\;\;\;\;
\begin{bmatrix}
\phantom{0} & \phantom{0} &   \\
 &  & \scriptstyle{0|2} \\
 &  & \scriptstyle{0|2}
\end{bmatrix}\!.
\end{equation}
We cannot have $A(U_2)=3$, since there is a contradiction in $\RR_1(T_3)$ and  
$C(T_i)=+$ for all tables. In addition, due to Rule~5, it is not possible that 
$A(U_2)=4$. Finally, we choose $a(U_2)=4,$ the other option would be 
$\alpha(U_2)=4$, this will be
discussed below.

{From} Observation~\ref{o28709} we can conclude that $\CC_1(T_1)$ and 
$\CC_1(T_4)$
do not contain contradictions, since $\CC_1(T_2)$ contains already a contradiction.
So $\CC_1(T_1)$ and $\CC_1(T_4)$ must differ in two places (Rule 5). 
One of these
places must be $A(T_1) \neq A(T_4)$.
Let us assume that the second one is $a(T_1) \neq a(T_4),$ the other case 
[$\alpha(T_1) \neq \alpha(T_4)$]
will be discussed below. Then, we can conclude that in $\RR_1(U_1)$ and 
$\CC_1(U_1)$ we cannot have the entries ``2'' and ``4'', and in $\RR_2(U_4)$ 
and $\CC_1(U_4)$ we cannot have the entries ``2'' and ``1''. To see this, note 
that we must have $A(T_2) = A(T_1) \neq A(T_4)$
and, if $B(U_1)= 2$, we can consider the measurement sequence $A_2 B_1 a_2 A_4$ 
or, if $B(U_1)= 4$, the sequence $A_2 B_1 A_4$ etc.  Hence, we have:
\begin{equation}\nonumber
U\!:\!
\begin{bmatrix}
{0} & {0} & {0}  \\
\scriptstyle{0|3} &  & 2 \\
\scriptstyle{0|3} &  & 3 
\end{bmatrix}\!,
\;\;
\begin{bmatrix}
1 & \phantom{0} & \scriptstyle{0|4}  \\
4 & \scriptstyle{0|4} & \scriptstyle{0|4} \\
 \scriptstyle{0|4}& \scriptstyle{0|4} & 0
\end{bmatrix}\!,
\;\;
\begin{bmatrix}
\phantom{0} & \phantom{0} & {0}  \\
0 & 0 & 0 \\
0 & 0 &  0
\end{bmatrix}\!,
\;\;
\begin{bmatrix}
\scriptstyle{0|3} & \phantom{0} &   \\
\scriptstyle{0|3} & \scriptstyle{0|3} & 0 \\
\scriptstyle{0|3} &  & \scriptstyle{0|2}
\end{bmatrix}\!.
\end{equation}
Here, we used in $\RR_1(U_1)$ that $\RR_1(T_3)$ has a contradiction and $C(T_i)=+$ 
for all tables, so it is not possible to go there.

Now, by Rule~1, we may fix $A(T_2) = a(T_2)= +.$ Then  we arrive at
\begin{equation}
T:
\begin{bmatrix}
+ & + & + \\
 &  & + \\
 &  & + 
\end{bmatrix}\!,
\;\;
\begin{bmatrix}
+ & + & + \\
+ & + & + \\
- & + & - 
\end{bmatrix}\!,
\;\;
\begin{bmatrix}
\phantom{+}& \phantom{-} & + \\
 & & - \\
 & & + 
\end{bmatrix}\!,
\;\;
\begin{bmatrix}
-& - & + \\
+ & & + \\
- & & - 
\end{bmatrix}\!,
\end{equation}
\begin{equation}
U:
\begin{bmatrix}
{0} & {0} & {0}  \\
\scriptstyle{0|3} &  & 2 \\
\scriptstyle{0|3} &  & 3 
\end{bmatrix}\!,
\;\;
\begin{bmatrix}
1 & \phantom{0} & \scriptstyle{0|4}  \\
4 & \scriptstyle{0|4} & \scriptstyle{0|4} \\
 \scriptstyle{0|4}& \scriptstyle{0|4} & 0
\end{bmatrix}\!,
\;\;
\begin{bmatrix}
\phantom{0} & \phantom{0} & {0}  \\
0 & 0 & 0 \\
0 & 0 &  0
\end{bmatrix}\!,
\;\;
\begin{bmatrix}
\scriptstyle{0|3} & \phantom{0} &  \scriptstyle{0|2} \\
0 & 0 & 0 \\
\scriptstyle{0|3} &  & \scriptstyle{0|2}
\end{bmatrix}\!.
\end{equation}
Here, we must have $A(T_4)=\alpha(T_4)$ since $\CC_1(T_4)$ has no 
contradiction.  Furthermore, $\RR_1(T_4)$ has no contradiction due to
Observation~\ref{o28709}. The values of $\RR_2(U_4)$ are determined by 
considering sequences like $c_4 a_4 c_?$; and $C(U_4) \neq 3$, because of 
$c_4^+ C_4 c_3^-$, and $C(U_4) \neq 1$, because of $c_4^+ C_4 \gamma_1 c_3^-.$

In addition, we can conclude that $A(U_4)=0$ and $B(U_4)=0$, since $\RR_1(T_3)$ 
has a contradiction and $C(T_i)=+$ for all tables, so it is not possible to go 
there. Then we can fill $T_4$  completely.  Then, also $C(U_4)=0,$ otherwise 
the sequence $B_4^- C_4^+ B_2^+$ gives a contradiction. If we had 
$\alpha(U_4)=3$, then we must have $A(T_4)=A(T_3)=-$ and, consequently (Rule~5) 
$B(U_3)=1$ or $2$, but then the sequence $A_4^- \alpha_4^- B_3^+ A_{1|2}^+$ 
leads to a contradiction, so $\alpha(U_4)=0.$ In summary, we have:
\begin{equation}
T:
\begin{bmatrix}
+ & + & + \\
 &  & + \\
 &  & + 
\end{bmatrix}\!,
\;\;
\begin{bmatrix}
+ & + & + \\
+ & + & + \\
- & + & - 
\end{bmatrix}\!,
\;\;
\begin{bmatrix}
\phantom{+}& \phantom{-} & + \\
 & & - \\
 & & + 
\end{bmatrix}\!,
\;\;
\begin{bmatrix}
-& - & + \\
+ & +& + \\
- & -& - 
\end{bmatrix}\!,
\end{equation}
\begin{equation}
U:
\begin{bmatrix}
{0} & {0} & {0}  \\
\scriptstyle{0|3} &  & 2 \\
\scriptstyle{0|3} &  & 3 
\end{bmatrix}\!,
\;\;
\begin{bmatrix}
1 & \phantom{0} & \scriptstyle{0|4}  \\
4 & \scriptstyle{0|4} & \scriptstyle{0|4} \\
 \scriptstyle{0|4}& \scriptstyle{0|4} & 0
\end{bmatrix}\!,
\;\;
\begin{bmatrix}
\phantom{0} & \phantom{0} & {0}  \\
0 & 0 & 0 \\
0 & 0 &  0
\end{bmatrix}\!,
\;\;
\begin{bmatrix}
0 & 0 &  0 \\
0 & 0 & 0 \\
0 &  & \scriptstyle{0|2}
\end{bmatrix}\!.
\end{equation}
Now $T_1$ is the only candidate for a table with three contradictions.  In 
order to obey Observation~\ref{o28709}, the only possibilities for 
contradictions are $\CC_2$, $\CC_3$, and $\RR_2$, since
$T_4$ has its contradiction in $\RR_3.$ Especially, there must be a contradiction in 
$\CC_2(T_1)$. Then, in order to obey Rule 5, we must have:
\begin{equation}
T:
\begin{bmatrix}
+ & + & + \\
 &  & + \\
 &  & + 
\end{bmatrix}\!,
\;\;
\begin{bmatrix}
+ & + & + \\
+ & + & + \\
- & + & - 
\end{bmatrix}\!,
\;\;
\begin{bmatrix}
-& + & + \\
 & & - \\
 & & + 
\end{bmatrix}\!,
\;\;
\begin{bmatrix}
-& - & + \\
+ & +& + \\
- & -& - 
\end{bmatrix}\!,
\end{equation}
\begin{equation}
U:
\begin{bmatrix}
{0} & {0} & {0}  \\
\scriptstyle{0|3} & \scriptstyle{2|3} & 2 \\
\scriptstyle{0|3} &  \scriptstyle{2|3}& 3 
\end{bmatrix}\!.
\;\;
\begin{bmatrix}
1 & \phantom{0} & \scriptstyle{0|4}  \\
4 & \scriptstyle{0|4} & \scriptstyle{0|4} \\
 \scriptstyle{0|4}& \scriptstyle{0|4} & 0
\end{bmatrix}\!,\!\;
\begin{bmatrix}
4 & \scriptstyle{1|2}  & {0}  \\
0 & 0 & 0 \\
0 & 0 &  0
\end{bmatrix}\!,\!\;
\begin{bmatrix}
0 & 0 &  0 \\
0 & 0 & 0 \\
0 &  & \scriptstyle{0|2}
\end{bmatrix}\!.
\end{equation}
However, if $b(U_1)=3$, then the sequence $B_1^+ b_1 A_3 B_4^-$
leads to a contradiction, while, if $\beta(U_1)=3$, then the sequence $B_1^+ 
\beta_1 A_3 B_4^-$ leads to a problem.

Finally, if we would have taken $\alpha(U_2)=4$ or $\alpha(T_1) \neq 
\alpha(T_4)$ the proof would proceed along the same lines, but this time the 
contradiction in $T_4$ would be in the second row.

{\bf Case 5B:} Let us assume that $\gamma(U_2)=4.$ Then, many entries on $U_4$ 
are fixed and we have:
\begin{equation}
T:
\begin{bmatrix}
\phantom{-} & \phantom{-} & + \\
 &  & + \\
 &  & + 
\end{bmatrix}\!,
\;\;
\begin{bmatrix}
\phantom{+} & \phantom{-} & + \\
 &  & + \\
 &  & - 
\end{bmatrix}\!,
\;\;
\begin{bmatrix}
\phantom{+}& \phantom{-} & + \\
 & & - \\
 & & + 
\end{bmatrix}\!,
\;\;
\begin{bmatrix}
\phantom{+}& \phantom{-} & + \\
 & & + \\
 & & - 
\end{bmatrix}\!,
\end{equation}
\begin{equation}
U:
\begin{bmatrix}
\phantom{0} & \phantom{0} & \phantom{0}  \\
 &  & 2 \\
 &  & 3 
\end{bmatrix}\!,
\;\;
\begin{bmatrix}
\phantom{0} & \phantom{0} & \scriptstyle{0|4}  \\
 \scriptstyle{0|4}& \scriptstyle{0|4} & \scriptstyle{0|4} \\
 &  & 4
\end{bmatrix}\!,
\;\;
\begin{bmatrix}
\phantom{0} & \phantom{0} & {0}  \\
0 & 0 & 0 \\
0 & 0 &  0
\end{bmatrix}\!,
\;\;
\begin{bmatrix}
\phantom{0} & \phantom{0} &  \scriptstyle{0|2} \\
 \scriptstyle{0|2}& \scriptstyle{0|2} & \scriptstyle{0|2} \\
\scriptstyle{0|2} & \scriptstyle{0|2} & \scriptstyle{0|2}
\end{bmatrix}\!.
\end{equation}
Here we cannot have  $a(U_4)=1$, due the sequences $c_2 \gamma_2 a_4 c_1$ [if 
$c(U_2)=0$] or $c_2 a_4 c_1$ [if $c(U_2)=4$], and also not $a(U_4)=3$, due to 
similar sequences. The same arguments apply to $b(U_4)$. The entries in 
$\RR_3(U_4)$  and $\CC_3(4)$
come from possible sequences like $\gamma_4 \alpha_4 \gamma_?$
if $\gamma(U_4)=0$ or $\gamma_4 \gamma_2 \alpha_4 \gamma_?$
if $\gamma(U_4)=2$.

But then the proof can proceed exactly as in the Case 5A, with 
$T_2$ and $T_4$ interchanged: The only significant difference 
comes from $c(U_1)=2 \neq 4,$ but this was never used in the proof.

\noindent
\\
{\bf Case 6:} {\it For $T_4$ one has $[C,c,\gamma]=[+-+]$:\\}
This is the same as the Case 5 with a permutation of $\RR_2$
and $\RR_3.$

\noindent
\\
{\bf Case 7:} {\it For $T_4$ one has $[C,c,\gamma]=[-++]$:\\}
In this case, the tables read:
\begin{equation}
T:
\begin{bmatrix}
\phantom{-} & \phantom{-} & + \\
 &  & + \\
 &  & + 
\end{bmatrix}\!,
\;\;
\begin{bmatrix}
\phantom{+} & \phantom{-} & + \\
 &  & + \\
 &  & - 
\end{bmatrix}\!,
\;\;
\begin{bmatrix}
\phantom{+}& \phantom{-} & + \\
 & & - \\
 & & + 
\end{bmatrix}\!,
\;\;
\begin{bmatrix}
\phantom{+}& \phantom{-} & - \\
 & & + \\
 & & + 
\end{bmatrix}\!,
\end{equation}
\begin{equation}
U:
\begin{bmatrix}
\phantom{0} & \phantom{0} & \phantom{0}  \\
 &  &  2\\
 &  &  3
\end{bmatrix}\!,
\;\;
\begin{bmatrix}
\phantom{0} & \phantom{0} & {0}  \\
\scriptstyle{0|4} & \scriptstyle{0|4} & 0 \\
 0& 0 & 0 
\end{bmatrix}\!,
\;\;
\begin{bmatrix}
\phantom{0} & \phantom{0} & {0}  \\
0 & 0 & 0 \\
 \scriptstyle{0|4}& \scriptstyle{0|4} &  0
\end{bmatrix}\!,
\;\;
\begin{bmatrix}
{0} & {0} & {0}  \\
\scriptstyle{0|2} & \scriptstyle{0|2} & 0 \\
\scriptstyle{0|3} & \scriptstyle{0|3} & 0 
\end{bmatrix}\!.
\end{equation}
Here, the entries in $U_1$ have been chosen without loss of generality: {From} 
Rule 4 and 5 it follows that one can restrict the attention to the cases where
$\CC_3(U_1)=[\;\;\;,2,3]$, 
$\CC_3(U_1)=[\;\;\;,2, 4]$, or
$\CC_3(U_1)=[\;\;\;,4,3]$. 
We only consider the first possibility, in the other cases the proof is 
analogous and is left to the gentle reader as an exercise.
The zeroes in $U_2, U_3$, and $U_4$ come from Rule~3. The entries  
$\scriptstyle{0|2}$ in $U_4$ come from possible measurement sequences like $c_4 
a_4 c_3$  or  $c_4 a_4 \gamma_1 c_3$ which prove that there cannot be the 
entries ``3'' or ``1''. The other entries can be derived accordingly.

{From} Rule~5, it follows that in $T_4$ the contradiction cannot be
in the rows, so it has to be in the first or second column. Let us assume, 
without loss of generality, that it is in $\CC_1(T_4)$.  Further, we can assume 
without loss of generality, that the values $A$ and $a$ in $T_4$ are both 
``$+$''.  Then, the tables can be more specified as \begin{equation}
T:
\begin{bmatrix}
\phantom{-} & \phantom{-} & + \\
 &  & + \\
 &  & + 
\end{bmatrix}\!,
\;\;
\begin{bmatrix}
{+} & \phantom{-} & + \\
 + & + & + \\
 + & - & - 
\end{bmatrix}\!,
\;\;
\begin{bmatrix}
{+}& \phantom{-} & + \\
 -&+ & - \\
 -& -& + 
\end{bmatrix}\!,
\;\;
\begin{bmatrix}
{+}& {-} & - \\
 + & + & + \\
 - & -& + 
\end{bmatrix}\!,
\end{equation}
\begin{equation}
U:
\begin{bmatrix}
\phantom{0} & \phantom{0} & \phantom{0}  \\
 &  &  2\\
 &  &  3
\end{bmatrix}\!,
\;\;
\begin{bmatrix}
\scriptstyle{0|1} & \phantom{0} & {0}  \\
0 & \scriptstyle{0|4} & 0 \\
 0& 0 & 0 
\end{bmatrix}\!,
\;\;
\begin{bmatrix}
\scriptstyle{0|1} & \phantom{0} & {0}  \\
0 & 0 & 0 \\
0& \scriptstyle{0|4} &  0
\end{bmatrix}\!,
\;\;
\begin{bmatrix}
{0} & {0} & {0}  \\
2 & \scriptstyle{0|2} & 0 \\
3 & \scriptstyle{0|3} & 0 
\end{bmatrix}\!.
\end{equation}
To see this, one first fills $T_4$, then, together with the entries of 
$\CC_1(U_4)$,
many values of $T_2$ and $T_3$ are fixed. The entries $\scriptstyle{0|1}$ are justified 
similar to the reasoning above. 

In $T_2$ as well as in $T_3$ the contradiction has to be either in $\RR_1$ or 
$\CC_2.$ However, there cannot be a contradiction in $\RR_1.$ To see this, 
assume that there were a contradiction in $\RR_1(T_2)$. 
Then, starting from $T_2$ we may measure the sequence $C_2 A_2 B$ or $C_2 B_2 A.$ 
According to Rule 5, we must end in two different $T_i.$ But the memory state can 
never change to $T_4$ [because $C(T_4)=-$]. So we must have
$B(U_2)=3,$ but this will not escape the contradiction, since
the values for $A$ and $C$ coincide in $T_2$ and $T_3$. 
So there is only $T_1$ left, and we arrive at a contradiction.

Consequently, the contradictions have to be both in $\CC_2(T_2)$ and  $\CC_2(T_3).$
In principle, our Observation~\ref{o28709} implies already that we cannot find 
a solution with three contradictions in one table. But one can also directly 
prove that there is no solution at all. We have:
\begin{equation}
T:
\begin{bmatrix}
+ & + & + \\
 & + & + \\
 & - & + 
\end{bmatrix}\!,
\;\;
\begin{bmatrix}
{+} & + & + \\
 + & + & + \\
 + & - & - 
\end{bmatrix}\!,
\;\;
\begin{bmatrix}
{+}& + & + \\
 -&+ & - \\
 -& -& + 
\end{bmatrix}\!,
\;\;
\begin{bmatrix}
{+}& {-} & - \\
 + & + & + \\
 - & -& + 
\end{bmatrix}\!,
\end{equation}
\begin{equation}
U:
\begin{bmatrix}
\phantom{0} & \phantom{0} & \phantom{0}  \\
 &  &  2\\
 &  &  3
\end{bmatrix}\!,
\;\;
\begin{bmatrix}
\scriptstyle{0|1} & 1 & {0}  \\
0 & 4 & 0 \\
 0& 0 & 0 
\end{bmatrix}\!,
\;\;
\begin{bmatrix}
\scriptstyle{0|1} & 1 & {0}  \\
0 & 0 & 0 \\
0& 4 &  0
\end{bmatrix}\!,
\;\;
\begin{bmatrix}
{0} & {0} & {0}  \\
2 & \scriptstyle{0|2} & 0 \\
3 & \scriptstyle{0|3} & 0 
\end{bmatrix}\!.
\end{equation}
Here, we must have $B(T_1)=B(T_2)=B(T_3)= +$ due to measurement 
sequences like $B_2^+B_1^+$ or $B_3^+ B_1^+$ and 
$\beta(T_1)=\beta(T_2)$ due to $\beta_2^- B_2^+ \beta_1^-$ and
$b(T_1)=b(T_3)$ due to $b_3^+ B_3^+ b_1^+.$ But then, starting from $T_2$, the 
sequence $\beta_2^- B_2^+ b_1^+$ reveals a contradiction to the PM conditions.

\noindent
\\
{\bf Case 8:} {\it For $T_4$ one has $[C,c,\gamma]=[---]$:\\}
In this case, we directly have:
\begin{equation}
T:
\begin{bmatrix}
\phantom{-} & \phantom{-} & + \\
 &  & + \\
 &  & + 
\end{bmatrix}\!,
\;\;
\begin{bmatrix}
\phantom{+} & \phantom{-} & + \\
 &  & + \\
 &  & - 
\end{bmatrix}\!,
\;\;
\begin{bmatrix}
\phantom{+}& \phantom{-} & + \\
 & & - \\
 & & + 
\end{bmatrix}\!,
\;\;
\begin{bmatrix}
\phantom{+}& \phantom{-} & - \\
 & & - \\
 & & - 
\end{bmatrix}\!,
\end{equation}
\begin{equation}
U:
\begin{bmatrix}
\phantom{0} & \phantom{0} & \phantom{0}  \\
 &  & 2 \\
 &  & 3 
\end{bmatrix}\!,
\;\;
\begin{bmatrix}
\phantom{0} & \phantom{0} & {0}  \\
0 & 0 & 0 \\
 0& 0 & 0 
\end{bmatrix}\!,
\;\;
\begin{bmatrix}
\phantom{0} & \phantom{0} & {0}  \\
0 & 0 & 0 \\
0 & 0 &  0
\end{bmatrix}\!,
\;\;
\begin{bmatrix}
{0} & {0} & {0}  \\
 &  & 0 \\
 &  & 0 
\end{bmatrix}\!.
\end{equation}
Starting from $T_2$ we may measure the sequence $C_2 A_2 B$ or $C_2 B_2 A.$
According to Rule 5, we must end in two different $T_i.$ But the memory state
can neither change to $T_4$ [because $C(T_4)=-$] nor to $T_3$ [as $\RR_1(T_3)$ 
contains a contradiction].  So there is only $T_1$ left, and we arrive at a 
contradiction.

In summary, by considering all eight different cases we have shown that no 
four-state solution exists in which one table has three contradictions.  This 
proves the claim.

\section{A 10-state automaton obeying all sequences}\label{X:apten}
In this Appendix we show an example of a 10-state automaton that obeys the set 
 of all sequences $\sqset_\mathrm{all}$.
For that, we define 10 eigenstates of two compatible observables.
We let $\ket{A^-B^+}$ be a quantum state with $A\ket{A^-B^+}= -\ket{A^-B^+}$ 
 and $B\ket{A^-B^+}= +\ket{A^-B^+}$.
In this fashion we define the 10 states
 $\ket{A^+B^+}$,
 $\ket{A^-B^+}$,
 $\ket{C^+c^+}$,
 $\ket{C^-c^+}$,
 $\ket{\gamma^+\beta^+}$,
 $\ket{\gamma^-\beta^+}$,
 $\ket{\alpha^+a^+}$,
 $\ket{\alpha^-a^+}$,
 $\ket{a^+b^+}$, and
 $\ket{B^+b^+}$.
Any measurement of an observable from the PM square projects with finite 
 probability any state of the set onto another state of the set.
If e.g.\ the automaton is in state $\ket{A^-B^+}$ and we measure $c$, QM 
 predicts a chance of 50\% to get the outcome $+1$ yielding the state 
 $\ket{C^-c^+}$, and a 50\% chance to obtain $-1$ and the state $\ket{C^-c^-}$.
The former state is in the set of the 10 states and hence our automaton would 
 return $+1$ and change to the state $\ket{C^-c^+}$.
We furthermore define that, if both states predicted by QM are in the set of 
 the 10 states, then we prefer the state corresponding to the output of $+1$.
Together with an arbitrary choice of the initial state, this completes the 
 definition of the automaton.
By construction, this automaton is deterministic and obeys 
 $\sqset_\mathrm{all}$.

\bibliographystyle{nature}
\bibliography{the}

\begin{thebibliography}{10}

\bibitem{Specker:1960DIE}
Specker, E.
\newblock {\em Dialectica}{ \bf 14}(2--3), 239--246 (1960).

\bibitem{Kochen:1967JMM}
Kochen, S. and Specker, E.~P.
\newblock {\em J. Math. Mech.}{ \bf 17}(1), 59--87 (1967).

\bibitem{Bell:1966RMP}
Bell, J.~S.
\newblock {\em Rev. Mod. Phys.}{ \bf 38}(3), 447--452 (1966).

\bibitem{Mermin:1990PRL}
Mermin, N.~D.
\newblock {\em Phys. Rev. Lett.}{ \bf 65}(27), 3373 (1990).

\bibitem{Heywood:1983FPH}
Heywood, P. and Redhead, M. L.~G.
\newblock {\em Found. Phys.}{ \bf 13}(5), 481--499 (1983).

\bibitem{BechmannP:2000PRL}
{Bechmann-Pasquinucci}, H. and Peres, A.
\newblock {\em Phys. Rev. Lett.}{ \bf 85}(15), 3313--3316 (2000).

\bibitem{Spekkens:2008PRL}
Spekkens, R.~W.
\newblock {\em Phys. Rev. Lett.}{ \bf 101}(2), 020401 (2008).

\bibitem{Cabello:2010PRL}
Cabello, A.
\newblock {\em Phys. Rev. Lett.}{ \bf 104}(22), 220401 (2010).

\bibitem{Aharon:2008PRA}
Aharon, N. and Vaidman, L.
\newblock {\em Phys. Rev. A}{ \bf 77}(5), 052310 (2008).

\bibitem{Svozil:2009PRA}
Svozil, K.
\newblock {\em Phys. Rev. A}{ \bf 79}(5), 054306 (2009).

\bibitem{Svozil:2009XXX}
Svozil, K.
\newblock CDMTCS Research Report Series 353,  MAR  (2009).

\bibitem{Horodecki:2010XXX}
Horodecki, K., Horodecki, M., Horodecki, P., Horodecki, R., Paw{\l}owski, M.,
  and Bourennane, M.
\newblock arXiv:1006.0468.

\bibitem{Bell:1982FOP}
Bell, J.~S.
\newblock {\em Found. Phys.}{ \bf 12}(10), 989--999 (1982).

\bibitem{LaCour:2009PRA}
La~Cour, B.~R.
\newblock {\em Phys. Rev. A}{ \bf 79}(1), 012102 (2009).

\bibitem{Khrennikov:2009}
Khrennikov, A.~Y.
\newblock {\em Contextual Approach to Quantum Formalism}.
\newblock Springer, {Berlin},  (2009).

\bibitem{Toner:2003PRL}
Toner, B.~F. and Bacon, D.
\newblock {\em Phys. Rev. Lett.}{ \bf 91}(18), 187904 (2003).

\bibitem{Pironio:2003PRA}
Pironio, S.
\newblock {\em Phys. Rev. A}{ \bf 68}(6), 062102 (2003).

\bibitem{Buhrman:2010RMP}
Buhrman, H., Cleve, R., Massar, S., and de~Wolf, R.
\newblock {\em Rev. Mod. Phys.}{ \bf 82}(1), 665 (2010).

\bibitem{Holevo:1973PIT}
Holevo, A.~S.
\newblock {\em Probl. Inf. Trans.}{ \bf 9}(3), 177--183 (1973).

\bibitem{Galvao:2003PRL}
Galv\~{a}o, E.~F. and Hardy, L.
\newblock {\em Phys. Rev. Lett.}{ \bf 90}(8), 087902 (2003).

\bibitem{Peres:1990PLA}
Peres, A.
\newblock {\em Phys. Lett. A}{ \bf 151}(3--4), 107--108 (1990).

\bibitem{Cabello:2008PRL}
Cabello, A.
\newblock {\em Phys. Rev. Lett.}{ \bf 101}(21), 210401 (2008).

\bibitem{Kirchmair:2009NAT}
Kirchmair, G. et~al.
\newblock {\em Nature (London)}{ \bf 460}(7254), 494--497 (2009).

\bibitem{Amselem:2009PRL}
Amselem, E., R{\aa}dmark, M., Bourennane, M., and Cabello, A.
\newblock {\em Phys. Rev. Lett.}{ \bf 103}(16), 160405 (2009).

\bibitem{Moussa:2010PRL}
Moussa, O., Ryan, C.~A., Cory, D.~G., and Laflamme, R.
\newblock {\em Phys. Rev. Lett.}{ \bf 104}(16), 160501 (2010).

\bibitem{Mealy:1955BST}
Mealy, G.~H.
\newblock {\em Bell Systems Technical J.}{ \bf 34}, 1045 (1955).

\bibitem{RothJr:2009}
Roth~Jr., C.~H.
\newblock {\em Fundamentals of Logic Design}.
\newblock Thomson, Stanford, CT,  (2009).

\bibitem{Guhne:2010PRA}
G\"{u}hne, O. et~al.
\newblock {\em Phys. Rev. A}{ \bf 81}(2), 022121 (2010).

\bibitem{Cabello:2010PRA}
Cabello, A.
\newblock {\em Phys. Rev. A}{ \bf 82}(3), 032110 (2010).

\bibitem{Harrigan:2007XXX}
Harrigan, N., Rudolph, T., and Aaronson, S.
\newblock arXiv:0709.1149.

\bibitem{Dakic:2008PRL}
Daki\'{c}, B., {\v S}uvakov, M., Paterek, T., and Brukner, {\v C}.
\newblock {\em Phys. Rev. Lett.}{ \bf 101}(19), 190402 (2008).

\bibitem{Vertesi:2009PRA}
V\'ertesi, T. and P\'al, K.~F.
\newblock {\em Phys. Rev. A}{ \bf 79}(4), 042106 (2009).

\bibitem{Brukner:2009FP}
Brukner, {\v C}. and Zeilinger, A.
\newblock {\em Found. Phys.}{ \bf 39}, 677--689 (2009).

\bibitem{Galvao:2009PRA}
Galv\~ao, E.~F.
\newblock {\em Phys. Rev. A}{ \bf 80}(2), 022106 (2009).

\bibitem{Gallego:2010PRL}
Gallego, R., Brunner, N., Hadley, C., and Ac\'\i{}n, A.
\newblock {\em Phys. Rev. Lett.}{ \bf 105}(23), 230501 (2010).

\bibitem{Brukner:2004XXX}
Brukner, {\v C}., Taylor, S., Cheung, S., and Vedral, V.
\newblock arXiv:quant-ph/0402127.

\bibitem{Brassard:2006PRL}
Brassard, G. et~al.
\newblock {\em Phys. Rev. Lett.}{ \bf 96}(25), 250401 (2006).

\bibitem{Pawlowski:2009NAT}
Paw{\l}owski, M. et~al.
\newblock {\em Nature (London)}{ \bf 461}(7267), 1101--1104 (2009).

\end{thebibliography}

\end{document}